\documentclass[conference]{IEEEtran}
\IEEEoverridecommandlockouts
\usepackage{cite}
\usepackage{amsmath,amssymb,amsfonts}
\usepackage{algorithmic}
\usepackage{graphicx}
\usepackage{textcomp}
\usepackage{xcolor}
\usepackage[most]{tcolorbox}
\usepackage{multirow}
\usepackage{balance}
\usepackage[normalem]{ulem}
\usepackage[hyphens]{url} 
\usepackage[colorlinks=true, allcolors=black]{hyperref}
\usepackage{booktabs}

\tcbset{
  rqbox/.style={
    colback=gray!10,  
    colframe=gray, 
    fonttitle=\bfseries,
    center title,
    title=RQ\theRQ\ Results,  
    boxrule=1pt,
    arc=4pt,
    left=1em,
    right=1em,
    top=1em,
    bottom=1em,
  }
}
\newcounter{RQ}
\newcommand{\RQBox}[2][]{%
  \stepcounter{RQ}
  \begin{tcolorbox}[rqbox, title={RQ\theRQ\ #1}]
    #2
  \end{tcolorbox}
}

\def\BibTeX{{\rm B\kern-.05em{\sc i\kern-.025em b}\kern-.08em
    T\kern-.1667em\lower.7ex\hbox{E}\kern-.125emX}}

\begin{document}

\title{On the need to perform comprehensive evaluations of automated program repair benchmarks: Sorald case study\\
}


\author{\IEEEauthorblockN{Sumudu Liyanage}
\IEEEauthorblockA{\textit{University of Otago} \\
Dunedin, New Zealand \\
bamsu520@student.otago.ac.nz}
\and
\IEEEauthorblockN{Sherlock A. Licorish}
\IEEEauthorblockA{\textit{University of Otago} \\
Dunedin, New Zealand \\
sherlock.licorish@otago.ac.nz}
\and
\IEEEauthorblockN{Markus Wagner}
\IEEEauthorblockA{\textit{Monash University} \\
Melbourne, Australia\\
markus.wagner@monash.edu}
\and
\IEEEauthorblockN{Stephen G. MacDonell}
\IEEEauthorblockA{\textit{Victoria University} \\
Wellington, New Zealand \\
stephen.macdonell@vuw.ac.nz}
}
\maketitle
\begin{abstract}
In supporting the development of high-quality software, especially necessary in the era of LLMs, automated program repair (APR) tools aim to improve code quality by automatically addressing violations detected by static analysis profilers. Previous research tends to evaluate APR tools only for their ability to clear violations, neglecting their potential introduction of new (sometimes severe) violations, changes to code functionality and degrading of code structure. There is thus a need for research to develop and assess comprehensive evaluation frameworks for APR tools. This study addresses this research gap, and evaluates Sorald (a state-of-the-art APR tool) as a proof of concept. Sorald’s effectiveness was evaluated in repairing 3,529 SonarQube violations across 30 rules within 2,393 Java code snippets extracted from Stack Overflow. Outcomes show that while Sorald fixes specific rule violations, it introduced 2,120 new faults (32 bugs, 2088 code smells), reduced code functional correctness---as evidenced by a 24\% unit test failure rate---and degraded code structure, demonstrating the utility of our framework. Findings emphasize the need for evaluation methodologies that capture the full spectrum of APR tool effects, including side effects, to ensure their safe and effective adoption.
\end{abstract}


\section{Introduction}
The delivery of high quality code has been a fundamental challenge in the software industry over many years. With the rise of large language models (LLMs), this challenge has intensified, as developers increasingly copy and paste LLM-generated code without thorough validation\cite{hamer2024just}. This practice can introduce hidden bugs and vulnerabilities, potentially leading to user-facing failures, financial losses, and reputation damage\cite{majdinasab2024assessing}. Software practitioners have used various techniques to overcome this challenge. Static code analysis is a common technique for finding code flaws before execution\cite{zampetti2017open}. With static code analysis, software defects can be identified early in the development cycle before affecting the users of the software system. However, addressing all the violations reported by static analysis tools can be time-consuming for software developers. As a result, developers may ignore certain warnings, potentially leaving defects unresolved\cite{imtiaz2019challenges}. Moreover, manual fixes are not always effective and may fail to fully address the underlying issue. Automated program repair (APR) tools were introduced to mitigate this issue. Pattern-Based Automatic Program Repair (PAR)\cite{kim2013automatic} was one of the earliest automated program repair mechanisms introduced. APR tools then evolved, inspired by techniques such as genetic programming\cite{yuan2018arja} and machine learning\cite{lutellier2020coconut}. Recent advancements in APR have explored the use of LLMs for automatically repairing software defects\cite{wadhwa2023frustrated}, albeit the depth of their evaluation is often questioned.

Evaluating APR tools under different conditions is crucial because it helps to identify the drawbacks of current mechanisms and inform improvements. Current studies have focused largely on APR tools' fixing capabilities\cite{etemadi2022sorald,yin2024thinkrepair,li2024hybrid}, neglecting their potential introduction of new violations, changes to code semantics, and degrading of code structure. To fill this gap, we introduce an expanded framework that analyzes the tendency of APR tools to introduce new errors and their impact on code functionality and structure, going beyond APR tools' fixing capability. We assess our framework on Sorald\cite{etemadi2022sorald}, a state-of-the-art Java code repairing tool, boasting performance that rivals LLMs\cite{wadhwa2023frustrated}. We also use real-world code from Stack Overflow for our evaluation, going beyond artificial/simulated benchmarks (e.g., Defects4J\cite{just2014defects4j}, QuixBugs\cite{lin2017quixbugs}).

Our contributions comprise our APR evaluation framework, evidence of how a high-performing APR tool performs against our framework, a new real-world dataset with test cases for future code analysis research, and recommendations for extending research on source code analysis and manipulation.

The remaining sections of this study are organized as follows. Section II presents the related work and research questions. Section III documents the methodology and Section IV the results. Section V discusses our results and their implications. Section VI considers threats, and finally, Section VII provides concluding remarks.


\section{Related Work}
\textbf{Code Quality:} Code quality has been a topic of concern for decades\cite{ndukwe2023have} given that faults/errors in software can propagate and have lasting effects on an industry. Lotter et al.~\cite{lotter2018code} showed that software developers tend to reuse code (including from Stack Overflow) without reviewing such artifacts for potential bugs and vulnerabilities. Research by Licorish and Nishatharan\cite{licorish2021contextual} demonstrated real-world security risks posed by online code. With the popularity of LLMs, developers are also increasingly using AI-generated code without assessing its quality\cite{licorish2025comparing}, event though such code may be more buggy than that written by humans\cite{chong2024artificial}. Issues may particularly be evident when parameter settings are not optimal\cite{arora2024optimizing}. Therefore, efforts aimed at ensuring code quality remain essential.

\textbf{Static Analysis for Enhancing Code Quality:} Static analysis is used by the software engineering community to detect software faults/errors\cite{gomes2009overview}. Tools like FindBugs, Checkstyle, PMD, and SonarQube use static analysis to flag potential programming errors, coding standard violations, syntax violations, and security vulnerabilities. With static code analysis, developers can identify potential threats early in the software development life cycle (SDLC)\cite{10371505}, leading to reduced cost for fixing defects later\cite{10.1145/3544902.3546233}. Research has shown that integrating static analysis into the software development pipeline significantly enhances software quality\cite{wadhams2024automating}. While these tools are effective in identifying issues, however, they largely rely on human developers to interpret and address these issues, which introduces the risk of inconsistent and delayed fixes.

\textbf{Static Analysis-Inspired Code Repair:} To fill the gap between software defect detection and repair, there are several efforts aimed at automating code repair. One of the earliest contributions in this space is PAR, introduced by Kim et al.~\cite{kim2013automatic}, which generates patches by applying fix patterns mined from existing human-written patches. Subsequently, semantics-based repair tools such as DirectFix\cite{mechtaev2015directfix} and tools using genetic programming\cite{yuan2018arja} were introduced to advance automated program repair. With the rise of machine learning, data-driven approaches have also been introduced for automatic code repair. Lutellier et al.~\cite{lutellier2020coconut} introduced CoCoNuT, which combines convolutional neural networks (CNNs) with a novel context-aware neural machine translation (NMT) architecture and ensemble learning to automatically generate bug-fixing patches across multiple programming languages by learning from past code edits and their contexts.

Recently, APR tools have been introduced that leverage LLMs for program repair. One such tool is RepairAgent\cite{bouzenia2403repairagent}, which uses an autonomous agent based on an LLM to fix software flaws. Sorald is also a recently introduced rule-based repair tool that automatically suggests patches for selected SonarQube rule violations in Java programs, where its fix rate is seen to be higher than that of LLMs\cite{wadhwa2023frustrated}.

\textbf{Evaluation Frameworks for APR Tools:} Evaluating program repair techniques is essential to ensure the delivery of high-quality code repairs, avoid bias and ensure reproducibility. It also helps to build more efficient tools, mitigating the drawbacks of current techniques. Current research\cite{etemadi2022sorald,yin2024thinkrepair,li2024hybrid} has primarily focused on evaluating tools based on their fixing capabilities---specifically, how many bugs they can fix and whether the fixes are successful. Yet, assessing whether APR tools introduce new errors, break original code functionality, or degrade code structure while performing repairs remain out of reach. It is also important to evaluate APR tools’ capability on different datasets to understand the (in)consistency of these techniques. An emerging trend involves evaluating tools on realistic, noisy code sources, such as Stack Overflow\cite{lic2022}. This is particularly necessary because existing studies\cite{li2024hybrid,bouzenia2403repairagent} evaluate tools on standard benchmarks like Defects4J\cite{just2014defects4j} and QuixBugs\cite{lin2017quixbugs}, which are not representative of code in the wild. This research fills this gap by developing a comprehensive evaluation framework. The framework goes beyond checking the fixing capability of APR tools by examining the new violations that are introduced, how repairs affect original code semantics/functionality, and how repairs degrade code structure. We use a Stack Overflow dataset\cite{meldrum2020understanding}, which contains real-world code samples, and evaluate Sorald as a case study given its previous performance. 

Having developed our framework, the following research questions guided our evaluation.

\textbf{RQ1: To what extent is Sorald effective at automatically repairing SonarQube rule violations in real-world code?}\\
Motivation: Most studies have evaluated Sorald on standard benchmark datasets\cite{etemadi2022sorald,wadhwa2023frustrated}. How Sorald performs on real-world code remains an open question. Outcomes from this RQ provide a baseline for subsequent analyses (for RQ2-4).

\textbf{RQ2: What types of new errors are introduced in the code after applying Sorald’s automated repairs?}\\
Motivation: Although many researchers have evaluated Sorald’s capability\cite{etemadi2022sorald,wadhwa2023frustrated}, previous work has not analyzed the tool's potential to introduce new errors. In fact, APR research does not address this issue, which is a gap in the domain.

\textbf{RQ3: Do Sorald’s repairs preserve the functional behavior of the original code?}\\
Motivation: Previous research has not examined how Sorald and other APR tools change the functional behavior of code they repair. This research question addresses this gap, in testing the utility of the proposed framework.

\textbf{RQ4: Does Sorald negatively affect the structural quality of the original code?}\\
Motivation: Prior research has shown that code structure is a strong indicator of code quality\cite{jiang2008comparing}. Code structure refers to how code is organized, including for instance, code cohesion, coupling, and complexity. Therefore, evaluating how Sorald (and other APR tools) affect code structure is important for understanding their utility and impact on code quality.

\section{Methodology}
\subsection{Study Details and Framework}
\begin{figure*}[htbp]
\centering
\includegraphics[width=1\textwidth]{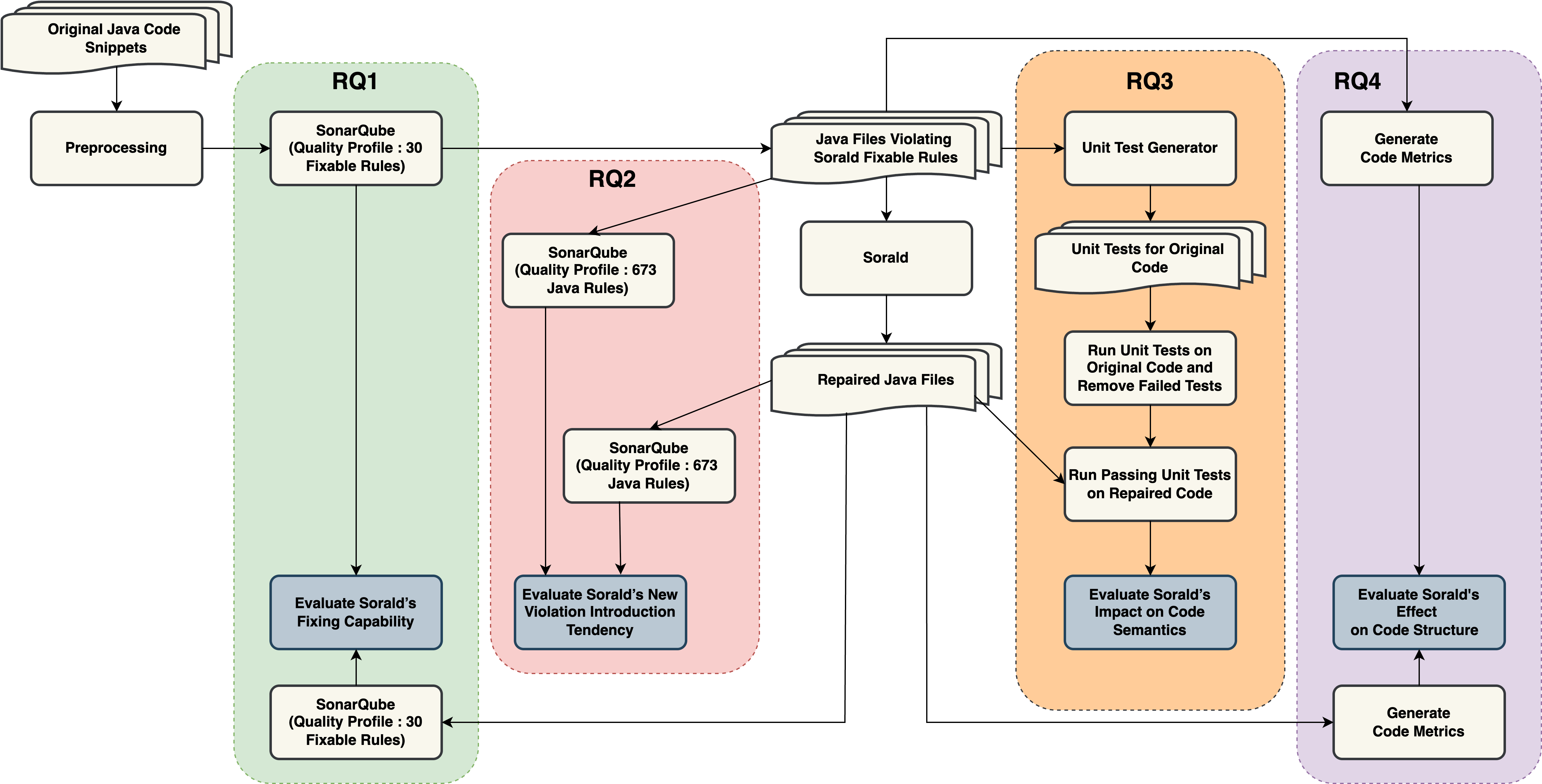}
\caption{Evaluation Framework and Experimental Design}
\label{figmethidology}
\end{figure*}

In evaluating the proposed framework (refer to Fig.~\ref{figmethidology}), our study seeks to evaluate Sorald's program repair capability on real-world code. Sorald is capable of fixing 30 types of SonarQube Java rule violations, including 18 types of bugs, 11 types of code smells, and 1 type of vulnerability (refer to Table~\ref{table:30rules}). 
Our framework is inspired by gaps in APR work\cite{etemadi2022sorald,yin2024thinkrepair,li2024hybrid}, the need to ascertain repair scope and code stability\cite{li2024hybrid,bouzenia2403repairagent} and the preservation of code structure\cite{jiang2008comparing} to support code quality. Thus, the key areas evaluated in this study include the extent to which Sorald can fix violations, the types of new violations introduced by Sorald, whether Sorald degrades the original code structure, and how Sorald affects the semantics/functionality of the original code. 

Fig.~\ref{figmethidology} depicts the framework, 
where it is noted that preprocessed snippets were analyzed by SonarQube using a quality profile configured with 30 fixable rules, in order to identify the .java files that violated these specific rules. Sorald was applied to the identified files to perform automated repairs, and SonarQube was run again using the same quality profile on the repaired code to determine how many violations were fixed by Sorald. The effect of Sorald on the original semantics of the code was analysed by generating unit tests for the original code and executing them on the repaired code to observe any test failures. The analysis of new violations introduced by Sorald during the repair process was performed by applying SonarQube using a quality profile configured with 673 Java rules for both the original and the fixed code, followed by a comparison of the resulting error reports. The impact on code structure was evaluated by comparing code metrics before and after the repair. For our framework in Fig.~\ref{figmethidology} to be generalizable, we could plug in any APR tool in place of Sorald and any static analysis tool in place of SonarQube.  

\begin{table*}[htbp]
\caption{30 SonarQube Java Rules Fixed By Sorald (In Application Order)}
\begin{center}
{\renewcommand{\arraystretch}{1.5}
\label{table:30rules}
\begin{tabular}{p{2cm}p{2.5cm}p{11cm}}
\toprule
\textbf{Rule} & 
\textbf{Type} & 
\textbf{Description} \\
\midrule
S1118&Code Smell&Utility classes should not have public constructors. \\

S1068&Code Smell&Unused ``private'' fields should be removed.\\

S1854&Code Smell&Unused assignments should be removed.\\

S1481&Code Smell&Unused local variables should be removed.\\

S1132&Code Smell&Strings literals should be placed on the left side when checking for equality.\\

S1444&Code Smell& ``public static'' fields should be constant.\\

S2184&Bug&Math operands should be cast before assignment.\\

S2142&Bug& ``InterruptedException'' should not be ignored.\\

S1948&Code Smell&Fields in a ``Serializable'' class should either be transient or serializable.\\

S2095&Bug&Resources should be closed.\\

S4973&Bug&Strings and Boxed types should be compared using ``equals()''.\\

S2057&Code Smell&Every class implementing Serializable should declare a static final serialVersionUID.\\

S2111&Bug& ``BigDecimal(double)'' should not be used.\\

S1656&Bug&Variables should not be self-assigned.\\

S2755&Vulnerability&XML parsers should not be vulnerable to XXE attacks.\\

S1155&Code Smell&Collection.isEmpty() should be used to test for emptiness.\\

S2116&Bug& ``hashCode'' and ``toString'' should not be called on array instances.\\

S1217&Bug& ``Thread.run()'' should not be called directly.\\

S2272&Bug& ``Iterator.next()'' methods should throw ``NoSuchElementException''.\\

S1860&Bug&Synchronization should not be based on Strings or boxed primitives.\\

S2097&Bug& ``equals(Object obj)'' should test argument type.\\

S3067&Bug& ``getClass'' should not be used for synchronization\\

S3984&Bug&Exception should not be created without being thrown.\\

S3032&Bug&JEE applications should not ``getClassLoader''.\\

S4065&Code Smell& ``ThreadLocal.withInitial'' should be preferred.\\

S2167&Bug& ``compareTo'' should not return ``Integer.MIN\_VALUE''.\\

S1596&Code Smell& ``Collections.EMPTY\_LIST", ``EMPTY\_MAP", and ``EMPTY\_SET'' should not be used.\\

S2204&Bug& ``.equals()'' should not be used to test the values of ``Atomic'' classes.\\

S2225&Bug& ``toString()'' and ``clone()'' methods should not return null.\\

S2164&Bug&Math should not be performed on floats.\\
\bottomrule
\end{tabular}
}
\label{tab1}
\end{center}
\end{table*}

\subsection{Dataset}
For this study, a dataset of 8010 Java code snippets previously extracted from Stack Overflow by Meldrum et al.\cite{meldrum2020understanding} was used. The original data set comprised Stack Overflow answers for questions with the Java tag from 2014 to 2016.
When we attempted to compile the 8010 Java files with Java JDK 17, 182 files could not be compiled due to differences in Java version compatibility. 
The remaining compilable 7828 Java files included files that did not violate any of Sorald's fixable 30 rules. Since those files were not useful for the study, we removed them from our dataset, resulting in 2393 files with violations that can be repaired by Sorald. Our replication package is available here\cite{liyanage_2025_16826817}.



\subsection{Evaluating Sorald's Fixing Capability (RQ1)}
We applied Sorald to the 2393 files, following the rule order specified in Table~\ref{table:30rules}. To assess Sorald’s code repair ability, we applied SonarQube, using a quality profile configured with the 30 targeted rules, to the Java files before repair. Then, SonarQube with the same quality profile was applied to the repaired Java files. Finally, the two error reports were compared to determine how many violations were solved.
The violation reports were initially retrieved in JSON format, and then we converted them into CSV format, in which each entry represented a violation. Two CSV files represented two states: before repair and after repair. Each entry in the CSV files included the file name, violated rule, start line, end line, severity, and violation type. These files were used to analyze the count of fixed violations and their distribution by violation type.
We did not simply compare the total violation counts to determine how many violations were resolved, because Sorald might have introduced new violations during repair, which can make the study imprecise. Instead, we performed a row-wise comparison for each entry in the pre-repair CSV file, checking whether a violation with the same {file name, rule, start line, end line} appeared in the post-repair CSV. If a match was found, it was discarded; if no match was found, the violation was considered fixed.

\subsection{Evaluating Sorald's New Violation Introduction (RQ2)}
To analyze new violations generated by Sorald, we first needed a suitable strategy to capture them. We obtained two CSV files, including the violations related to 673 SonarQube Java rules for Sorald's pre-repair and post-repair states. 
Sorald significantly changed the formatting of the original code which resulted in line position shift in the code. Thus, entries in the post-repair violation CSV that did not have an exact match in the pre-repair violation CSV, based on the file name, rule, start line, and end line of the violation, could not be considered as \emph{new} errors without close examination.

We identified the relevant code fragment associated with errors in the fixed code using the start line and end line details given in the violation entry in the post-repair violation CSV. We then checked whether this exact code fragment appeared in the same file in the original code. If it did, the violation was not considered new, as Sorald had not altered the code. If the fragment did not exist in the original file, we applied a final check where we looked for an exact match of the file name, rule, start line, and end line in the pre-repair violation CSV. If no such match was found, the violation was classified as a new error introduced by Sorald. This strategy was applied to each entry in the post-repair violation CSV.

Our above algorithm identified 2120 violations as newly introduced by Sorald. We used a binomial test approximation to prove the algorithm is accurate. To achieve this, we initially selected a sample from the newly identified violations. The sample size was calculated as 326 with a 95\% confidence level, a 5\% margin of error, and a 50\% population proportion, following established practices in software engineering research\cite{nguyen2023software}. We used the stratified sampling approach with proportional allocation, with an extra constraint: every violation rule must appear at least once. We obtained 359 samples from this criterion and adjusted the sample size to 326, reducing the sample counts for the most frequent rules while never reducing the count of any rule below 1. Afterwards, we manually verified them to identify false positives. This verification process was carried out by two evaluators for reliability. We verified that 250 out of 326 were true positives and 76 were false positives. We obtained a p-value of 0.0043 from the binomial test, using 70\% as the intended precision threshold, which is commonly considered practical\cite{jimenez2016vulnerability}. Given this p-value, we have statistically significant evidence to conclude that the true precision of our algorithm is greater than 70\%. We used the same sample to analyze the types of new violations and to see how critical they are. We also manually checked the reasons for these new errors by investigating the original code and the repaired code.

\subsection{Evaluating Sorald’s Impact on Code Semantics (RQ3)}
We checked the effect of Sorald on the functionality of the original code using unit tests. We generated unit tests for the 2393 original Java files using EvoSuite\cite{fraser2011evosuite}. The total number of tests generated was 8274. The generated test suite achieved an average coverage of 89\%, targeting a total of 135,960 combined coverage goals---including lines, branches, exceptions, mutation points, methods, and so on---and successfully covered 108,605 of them. These generated tests were first run on the original code. Afterwards, the test cases that failed on the original code were removed. The number of passed test cases was 8,212 of 8,274, covering 2,422 classes. These passing test cases on the original code were then executed on the repaired code. We then examined how many test cases failed, and we manually investigated the reasons for those failures, which revealed the effect of Sorald on the semantics of the code.

\subsection{Evaluating Sorald's Impact on Code Structure (RQ4)}
\begin{figure}[htbp]
\centering
\includegraphics[width=0.5\textwidth]{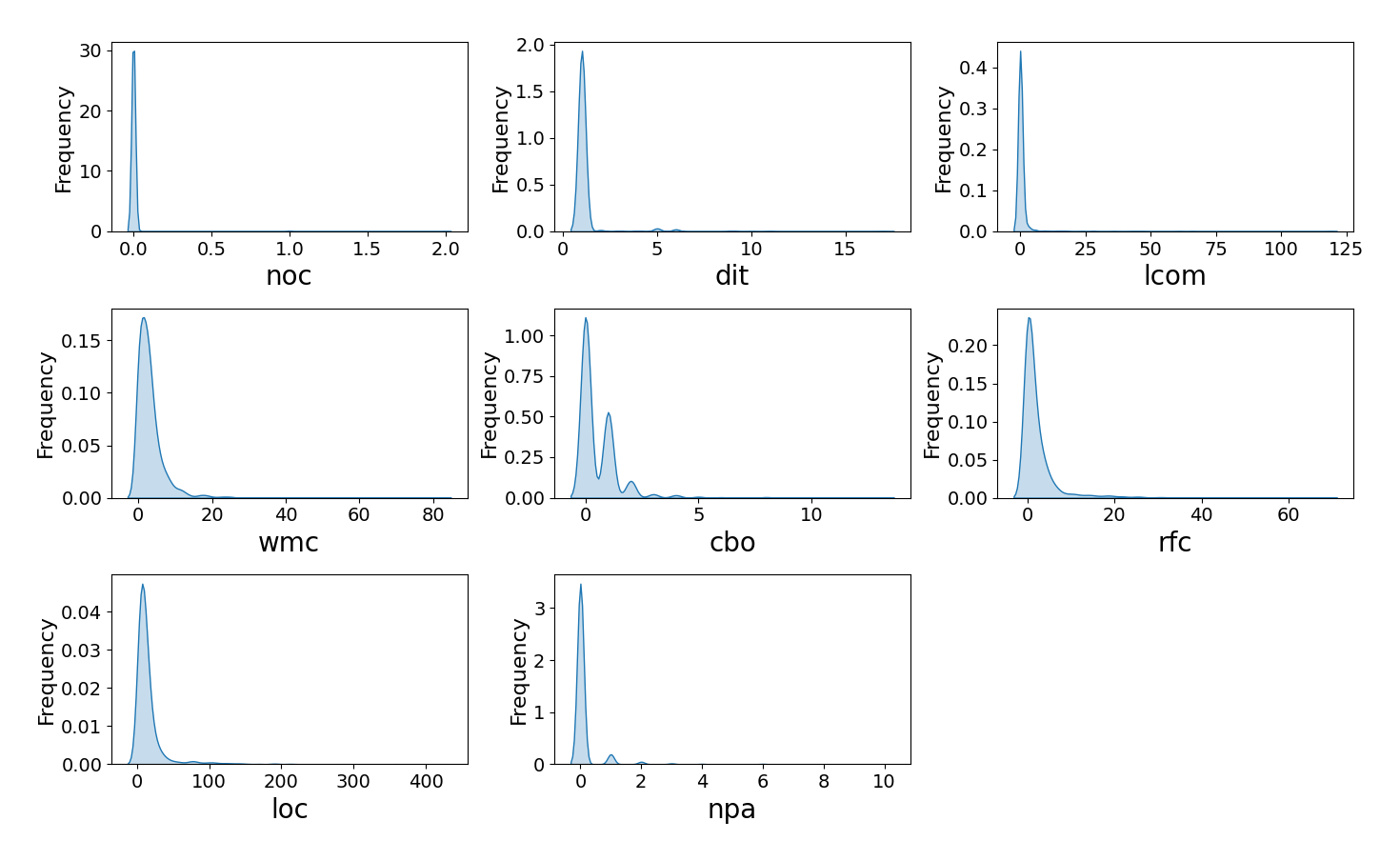}
\caption{Distribution of Code Metric Values Before Repair}
\label{figdistribution}
\end{figure}

It was important to investigate how Sorald affected the original code structure. For this analysis, we selected a set of code metrics that are proven to be linked to software defects. By examining changes in these metrics, we can investigate how Sorald has affected code quality. Several studies have been conducted to predict software defects using code metrics. In research conducted by Dominik et al.~\cite{rebro2023source}, NOC {(Number of Children)}, NPA {(Number of Public Attributes)}, DIT {(Depth of Inheritance Tree)}, and LCOM5 {(Lack of Cohesion of Methods, version 5)} were the highest-performing individual metrics for fault prediction. The LOC {(Lines of Code)} metric was suggested by Gyimothy et al.~\cite{gyimothy2005empirical} to be a suitable defect predictor. Many studies related to software defect prediction have also used the CK (Chidamber and Kemerer) metrics set\cite{chidamber1994metrics}---WMC {(Weighted Methods per Class)}, DIT, NOC, CBO {(Coupling Between Objects)}, RFC {(Response For a Class)}, and LCOM1\cite{gyimothy2005empirical}. In evaluating these studies, we selected NOC, NPA, DIT, LCOM1, WMC, CBO, RFC, and LOC given their repeated effects in previous work. All of these metrics were calculated by the CK tool\cite{aniche-ck}. The CK tool generated class-level metrics, and we aggregated them at the file level. NOC, NPA, LCOM1, WMC, CBO, RFC, and LOC were summed across all classes in the file, while DIT was taken as the maximum value among the classes in the file. We generated relevant code metrics for the pre-repair and post-repair code. Based on the data distribution (refer to Fig.~\ref{figdistribution}) and results of a D'Agostino-Pearson test, we confirmed that the data violated normality. Therefore, we performed the Wilcoxon signed rank test to identify which metrics changed significantly after the repair process. To understand the direction of change (increase or decrease), we used the median and mean signed ranks.

\section{Results}
\subsection{Evaluating Sorald’s Fixing Capability (RQ1)}
The original 2393 Java code snippet files had 3529 code violations as identified by SonarQube, configured with a quality profile of 30 Sorald fixable rules. Sorald  was able to remove 3423 of these violations, achieving a fix rate of 97.0\% for the targeted rule violations it was built to repair.
Table II shows the violations that existed in the original code and the corresponding percentage of violations that Sorald was able to fix. The violations spanned 21 different rule types. Rule S1118, which is a code smell indicating that utility classes should not have public constructors, had the highest number of violations (1,684 instances), and Sorald was able to fix 99.9\%. Sorald fixed this issue by introducing an empty private constructor at the top of the class. The next most frequent violations were S1068 (509 instances), S1854 (288 instances), and S1481 (281 instances), which were warnings to remove unused variables and assignments. Sorald fixed these violations by removing them, and the fix rates were 98.2\%, 99.3\%, and 100\%, respectively. Rule S1444, “public static fields should be constant,” occurred 101 times in the original code, and Sorald fixed this by making public static fields final. Rules S4973, S2057, S2111, S1656, S2755, S1155, S2116, S1217, S1860, and S2272 had the least number of violations (fewer than 25 each), and Sorald was able to fix all of them. Sorald showed the lowest fixing rates for rules S2164 and S1948, at 36.3\% and 45.3\%, respectively.

\begin{table}[h!]
\caption{Fix Rate of Sorald by Rule Type}
\renewcommand{\arraystretch}{1.2}
\centering
\begin{tabular}{cccc}
\toprule
\textbf{} & \textbf{Rule} & \textbf{Violation Count} & \textbf{Fixed Percentage} \\
\midrule
1&S1118 & 1684 & 99.9\% \\

2&S1068 & 509 & 98.2\% \\

3&S1854 & 288 & 99.3\% \\

4&S1481 & 281 & 100\% \\

5&S1132 & 212 & 100\% \\

6&S1444 & 101 & 100\% \\

7&S2184 & 90 & 100\% \\

8&S2142 & 85 & 100\% \\

9&S2164 & 80 & 36.3\% \\

10&S1948 & 75 & 45.3\% \\

11&S2095 & 46 & 95.7\% \\

12&S4973 & 24 & 100\% \\

13&S2057 & 14 & 100\% \\

14&S2111 & 11 & 100\% \\

15&S1656 & 9 & 100\% \\

16&S2755 & 7 & 100\% \\

17&S1155 & 5 & 100\% \\

18&S2116 & 4 & 100\% \\

19&S1217 & 2 & 100\% \\

20&S1860 & 1 & 100\% \\

21&S2272 & 1 & 100\% \\
\bottomrule
\end{tabular}
\label{tab:simple-table}
\end{table}

\RQBox[Results]{
Sorald was able to remove 97\% of the violations overall, eliminating 3423 out of 3529 violations. However, rules S2164 and S1948 had the lowest fix rates, at 36.3\% and 45.3\% respectively, while other Sorald-fixable rules achieved fix rates greater than 95\%.}

\subsection{Evaluating Sorald’s New Violation Introduction (RQ2)}
\begin{figure}[htbp]
\centering
\includegraphics[width=0.5\textwidth]{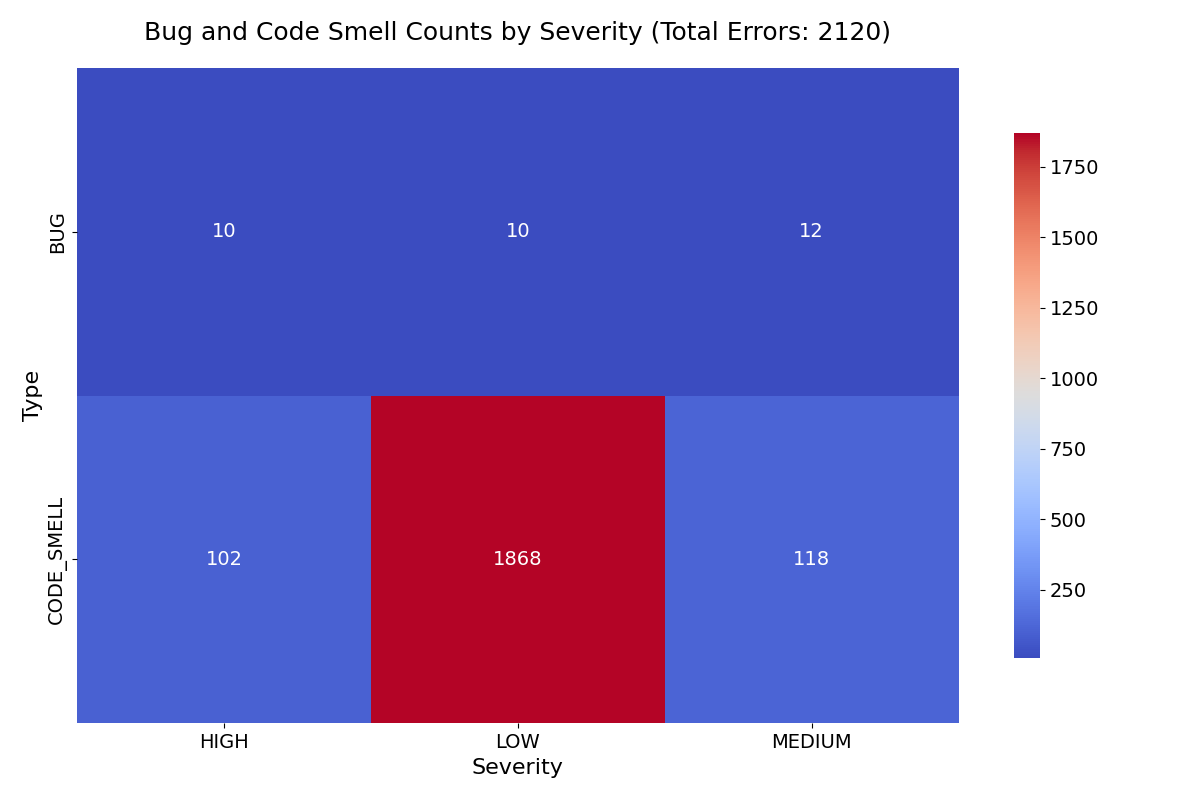}
\caption{Violations Introduced by Sorald, Categorized by Type and Severity}
\label{fignewerrorheatmap}
\end{figure}

\begin{figure}[htbp]
\centering
\includegraphics[width=0.5\textwidth]{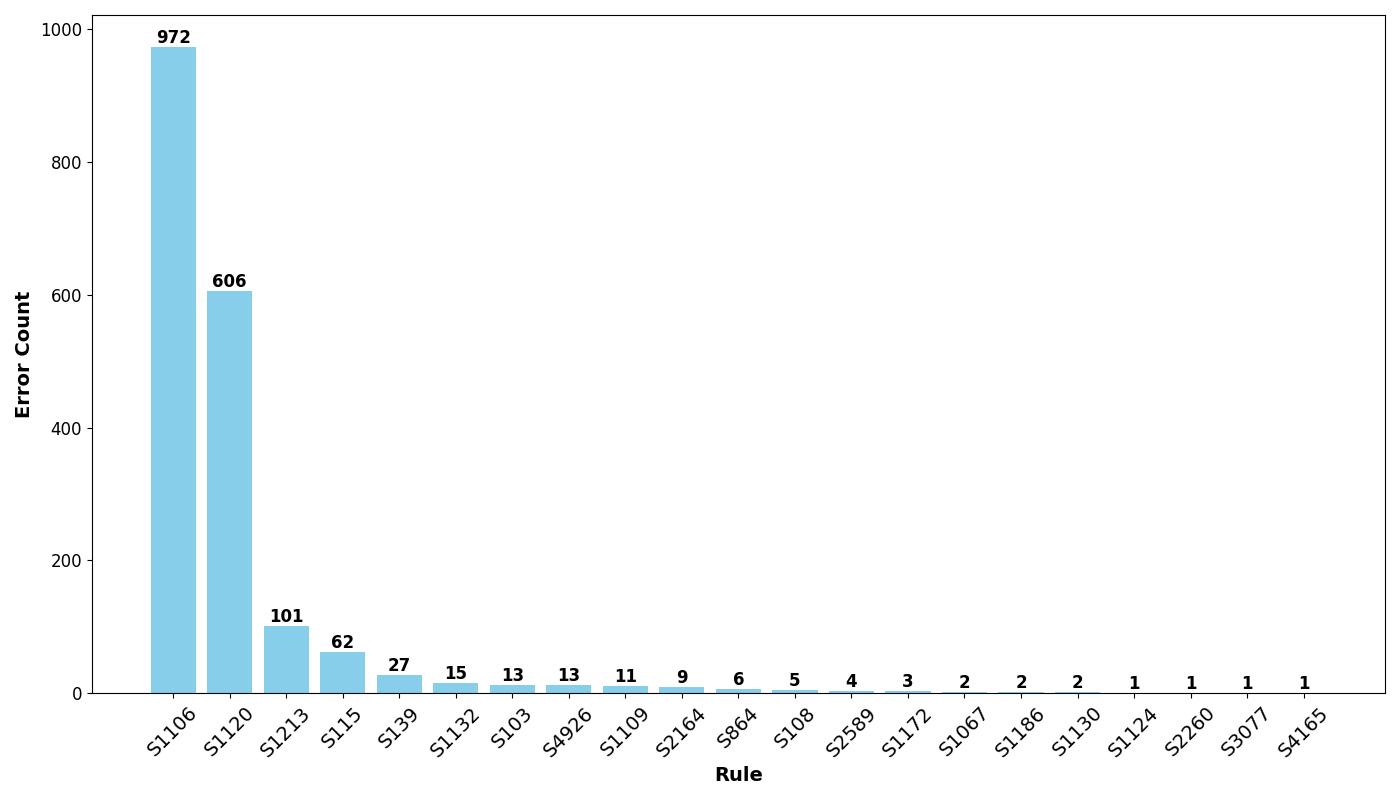}
\caption{Frequency of Violations Introduced by Sorald (Ordered by Frequency)}
\label{fignewerrorfrquency}
\end{figure}

After analyzing the original code and the repaired code with SonarQube using the 673 Java rules quality profile, 2120 violations were identified as new errors introduced by Sorald. SonarQube classifies violations into three types: vulnerabilities, bugs, and code smells, each of which is assigned a severity level of high, medium, or low. Fig.~\ref{fignewerrorheatmap} shows the distribution of introduced new errors across these categories and their severity, where it is observed that new vulnerabilities were not introduced by Sorald. However, Sorald introduced several new bugs and a large number of code smells while repairing violations. As noted above, we manually analyzed a sample of 326 of these 2120 errors to study these new errors. Table~\ref{tab:newerrors} shows violation types that were confirmed to be introduced by Sorald, and Fig.~\ref{fignewerrorfrquency} shows the frequency of these violations.

One of the bugs introduced---S2164 `Math should not be performed on floats'---can affect the precision of arithmetic computations. This error was introduced as a side effect of the fix for S2184 `Math operands should be cast before assignment’. To resolve S2184, Sorald appended the literal suffix `f' to the first operand of the expression when the result was a float. This modification caused the expression to be evaluated using float precision, however, thereby introducing S2164. Another bug introduced was S3077 `Use a thread-safe type; adding volatile is not enough to make this field thread-safe', which may lead to race conditions in concurrent environments. This occurred as a result of Sorald’s attempt to fix rule S1860 `Synchronization should not be based on Strings or boxed primitives.' To fix S1860, if the lock is a field of the current class where the synchronization block is located, then Sorald adds a new field as an Object lock. In this particular case, although Sorald is supposed to add this field as final, it mistakenly added it as volatile, introducing S3077.

Sorald introduced syntax errors into the original code, making it no longer parsable. This particular error was generated while Sorald was trying to remove unused assignments. Sorald also violated the original code’s coding standards by breaking naming conventions, converting expressions into more complex forms, declaring modifiers in the wrong order, disrupting the predefined order of class members, and introducing redundant assignments. This is because Sorald injects pieces of code into the original code or alters the existing code without fully considering all possible outcomes. Sorald also blindly introduced a serialVersionUID as 1L for all the classes that implement Serializable for object serialization to fix S2057. That fix, however, introduced S4926 `serialVersionUID should not be declared blindly,' as 1L might not work in some scenarios. Further, Sorald changed the original formatting of the code by introducing indentation issues, altering the location of comments, and violating standard conventions for the placement of curly braces. The repaired code was also left with empty methods and unused method parameters.

\RQBox[Results]{
Sorald introduced 2120 new violations in the code repair process, including several bugs and numerous code smells.}

\begin{table*}[htbp]
\caption{New Violations Introduced by Sorald (Ordered by Frequency)}
\begin{center}
{\renewcommand{\arraystretch}{1.5}
\label{table:newerrors}
\begin{tabular}{p{1cm}p{1.5cm}p{11cm}p{1.5cm}}
\toprule
\multicolumn{1}{c}{\textbf{Rule}} & 
\multicolumn{1}{c}{\textbf{Type}} & 
\multicolumn{1}{c}{\textbf{Description}}&
\multicolumn{1}{c}{\textbf{Severity}}\\
\midrule

S1106&Code Smell&An open curly brace should be located at the beginning of a line.&LOW\\

S1120&Code Smell&Source code should be indented consistently.&LOW\\

S1213&Code Smell&The members of an interface or class declaration should appear in a pre-defined
order.&LOW\\

S115&Code Smell& Constant names should comply with a naming convention.&HIGH\\

S139&Code Smell&Comments should not be located at the end of lines of code.&LOW\\

S1132&Code Smell&Strings literals should be placed on the left side when checking for equality.&LOW\\

S103&Code Smell&Lines should not be too long.&MEDIUM\\

S4926&Code Smell&serialVersionUID should not be declared blindly.&LOW\\

S1109&Code Smell&A close curly brace should be located at the beginning of a line.&LOW\\

S2164&Bug&Math should not be performed on floats.&LOW \\

S864&Code Smell&Limited dependence should be placed on operator precedence.&MEDIUM\\

S108&Code Smell&Nested blocks of code should not be left empty.&MEDIUM\\

S2589&Code Smell&Boolean expressions should not be gratuitous.&MEDIUM\\

S1172&Code Smell&Unused method parameters should be removed.&MEDIUM\\

S1067&Code Smell&Expressions should not be too complex.&HIGH\\

S1186&Code Smell&Methods should not be empty.&HIGH\\

S1130&Code Smell&Exceptions in ``throws'' clauses should not be superfluous.&LOW\\

S1124&Code Smell&Modifiers should be declared in the correct order.&LOW\\

S2260&Code Smell&Parse error.&MEDIUM\\

S3077&Bug&Use a thread-safe type; adding \"volatile\'' is not enough to make this field.&LOW\\

S4165&Code Smell&Assignments should not be redundant.&MEDIUM\\
\bottomrule
\end{tabular}
}
\label{tab:newerrors}
\end{center}
\end{table*}

\subsection{Evaluating Sorald’s Impact on Code Semantics (RQ3)}
\begin{figure}[htbp]
\centering
\includegraphics[width=0.5\textwidth]{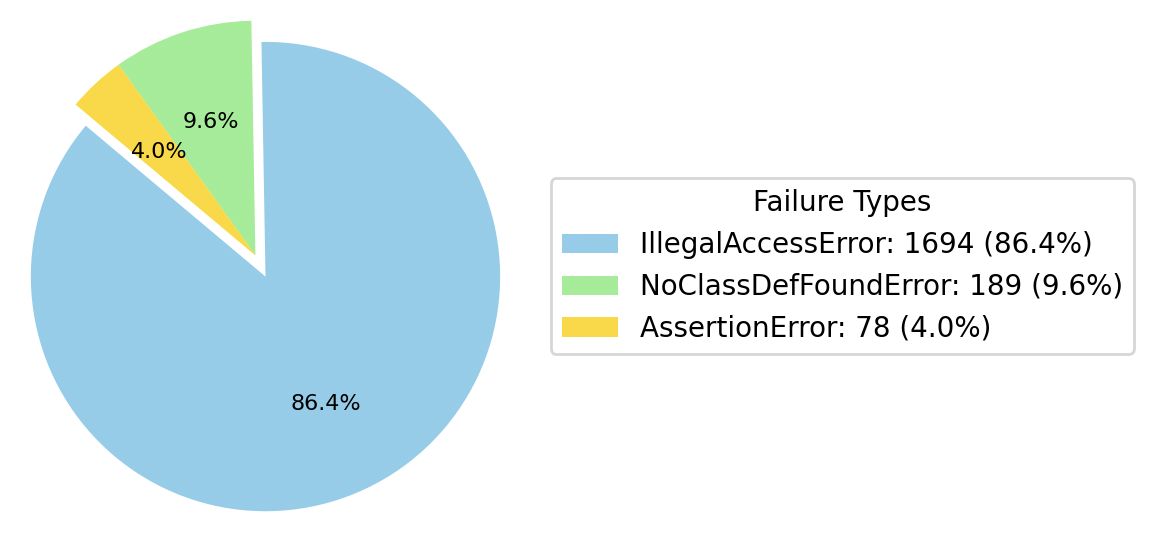}
\caption{Distribution of Test Failures by Failure Type}
\label{figuretestfailuredistribution}
\end{figure}

A total of 8212 test cases, all of which passed on the original code, were executed on the repaired code to assess Sorald's impact on code functionality. Of these, 1,962 tests failed on the repaired version, resulting in a passing rate of 76.1\%. Of the total failures, 1694 tests failed with IllegalAccessError, meaning tests were trying to access private methods in the code, 189 failed with NoClassDefFoundError, being unable to find the .class files of some Java code snippets, and 78 of the failures were due to assertion errors. One failure was caused by a simulation error and was excluded from the analysis. The distribution of failure types across the 1,961 failures is shown in Fig.~\ref{figuretestfailuredistribution}.

The reason for test failures with NoClassDefFoundError was that 61 files could not be compiled after applying Sorald. Twenty-four of these files failed to compile, saying “cannot find symbol”, typically indicating that Sorald removed or altered a code element that was still referenced elsewhere. Another 17 files failed to compile due to uninitialized variables, with the compiler reporting ``variable might not have been initialized.'' Moreover, 11 files produced the error “cannot assign a value to a final variable.” Among the less frequent compile errors, “not a statement,” “illegal combination of modifiers,” and “particular exception is never thrown in the body of corresponding try statement” could be seen. All compilation errors and their reasons are shown in Table IV.

We observed that 78 test cases failed because of assertion errors, with 59 of these failures occurring due to test cases expecting certain exceptions which were no longer thrown from the code after Sorald's repairs. The most frequently expected but unthrown exceptions included NullPointerException, FileNotFoundException, HeadlessException, ArithmeticException, NumberFormatException, and IOException. One of the main reasons for assertion failures was that Sorald had removed the entire original code instead of fixing it. Most of the other assertion failures were caused by Sorald's changes, such as removing unused variables, casting operands to float or double to fix S2184, and modifying string equality checks to fix S1132 by placing string literals on the left side. Most importantly, some of the changes Sorald made to arithmetic expressions led to incorrect outputs, which is a critical concern. The most frequent test failure type was IllegalAccessError. The reason for this was Sorald introduced a private constructor for each class that violated S1444, which led to the failure of all test cases that tried to create instances of these classes.

\begin{table*}[h]
\centering
{\renewcommand{\arraystretch}{1.5}
\label{tab:compileer}
\caption{Compile Errors Introduced By Sorald}
\begin{tabular}{p{4.5cm}p{12.5cm}}
  \toprule
  \textbf{Compile Error} & \textbf{Reason} \\ \midrule
  
  \multirow{7}{*}{Cannot find symbol} & Sorald randomly removed some necessary import statements in certain files. \\
  & When removing unused local variables (S1481), Sorald sometimes removed only the variable declarations but did not update all places where those variables were referenced. \\
  &Sorald rearranged certain lines of code to fix some SonarQube rules. As a result, some lines using variables were moved above their variable declarations.\\
  &Sorald mistakenly removed some variables that were still in use in certain cases.\\
  &In some cases, Sorald removed mistakenly the space between modifiers and variables where variables were defined, which caused the variable names to change.\\
  
  Variable might not have been initialized& To fix S1444 (“public static” fields should be constant), Sorald added the final modifier to public static variables without checking if they were initialized. As a result, it made even uninitialized variables final.\\
  
  Cannot assign a value to final variable&In some cases, public static fields were assigned values inside the main method. To fix S1444, Sorald blindly made them final, even though these variables were assigned values within methods.\\
  
  Not a statement&Sorald partially removed parts of some statements, leaving behind incomplete and meaningless expressions.\\
  
  Some exceptions are never thrown in body of corresponding try statement &Sorald removed some statements with unused variables but did not remove the related catch blocks for exceptions those statements could have thrown.\\
  
  Illegal combination of modifiers: final and volatile&While adding new modifiers to fix certain issues, Sorald ignored the original modifiers on the expressions. \\
  
  The parent class has private access&To fix S1118 (Utility classes should not have public constructors), Sorald had blindly introduced a private constructor for each class. So, child classes which extended those classes were no be compiled because of this change.\\
  \bottomrule
\end{tabular}
}
\label{tab:compileer}
\end{table*}

\RQBox[Results]{
For Sorald's repairs, 1962 test failures were observed for 8212 tests, with a passing rate of 76.1\%. The test failures were mostly due to Sorald changing the accessibility of classes and compilation errors. Moreover, several assertion errors were seen.}

\subsection{Evaluating Sorald's Impact on Code Structure (RQ4)}
\begin{figure}[htbp]
\centering
\includegraphics[width=0.5\textwidth]{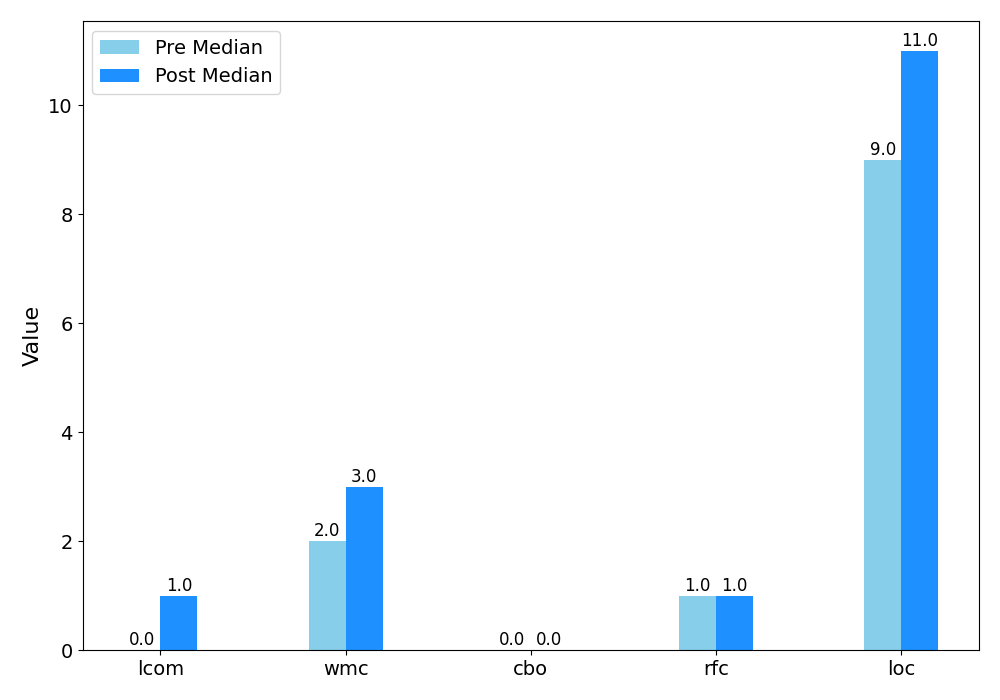}
\caption{Median Metric Values: Pre vs Post Repair}
\label{figuresppvalue}
\end{figure}

\begin{figure}[htbp]
\centering
\includegraphics[width=0.5\textwidth]{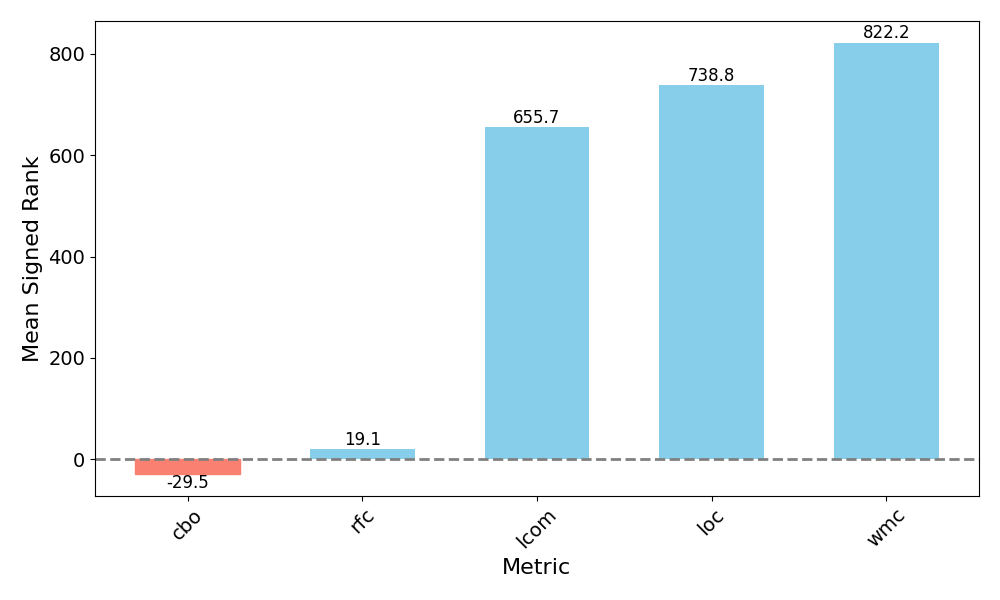}
\caption{Mean Signed Ranks of Metric Deltas (Post-Pre)}
\label{figurescoef}
\end{figure}

The Wilcoxon signed-rank test returned p-values of 0.000 for all metrics except DIT, NPA and NOC (values being 0.3173, 0.0633 and undefined-nan respectively). These results indicate that Sorald’s changes significantly affected LCOM1, WMC, CBO, RFC, and LOC. 
Based on the median and mean signed ranks of these significant metrics (Fig.~\ref{figuresppvalue}, Fig.~\ref{figurescoef}), it is observed that Sorald's repairs resulted in increased LCOM1, LOC, WMC and RFC. Higher values of these metrics suggest that Sorald has negatively affected the code structure. However, Sorald has slightly decreased CBO, a positive effect. 
\RQBox[Results]{
Sorald repairs negatively affected code structure, resulting in notably higher values for metrics LCOM1, LOC, WMC and RFC. However, CBO  decreased after Sorald's repairs, a positive effect. Overall, Sorald repairs degraded code structure.}


\section{Discussion and Implications}
We demonstrate the utility of our framework in answering the four research questions and discussing the implications of our outcomes in turn below.

\subsection{Evaluating Sorald’s Fixing Capability (RQ1)}
In evaluating the fixing capability of Sorald we found that the tool achieved an overall fix rate of 97\% for the 30 SonarQube violations, supporting prior outcomes\cite{etemadi2022sorald}. 
This demonstrates the tool's continued improvement and expanded capabilities. Although Sorald achieved a fix rate above 95\% for most of the rules, it failed to fix rules S2164 and S1948 effectively, being unable to fix at least 50\% of the occurrences of those violations.
Overall, Sorald achieved a high fixing rate because it was able to make the SonarQube warnings go away. However, Sorald can also act on false positives suggested by SonarQube, altering healthy code into buggy code.
For example, the most frequent violation we witnessed in our code was S1118 'utility classes should not have public constructors'. To fix this, Sorald introduced an empty private constructor to each class. It is true that this fix can take the SonarQube violation away, but at the same time, introducing a private constructor to a class without knowing the context can affect classes that extend this particular class or any location where this class is used. Sorald blindly applied fixes for the violations SonarQube was suggesting without knowing the context of the codebase.\\
\textbf{Implication:} Sorald can be considered a good tool to fix certain types of errors where the fix is straightforward and does not affect the rest of the code. However, even for straightforward fixes there is a threat, because Sorald may introduce bugs into healthy code, relying on false positives suggested by SonarQube. Sorald lacks context-awareness during repairs, and thus manual reviews are necessary after its application.

\subsection{Evaluating Sorald’s New Violation Introduction (RQ2)}
Sorald introduced a large number of code smells and several bugs during the repair process. We found that some specific errors could lead to issues such as precision loss in arithmetic expressions, the introduction of race conditions, and making the original code unparsable. Importantly, Sorald violated coding standards by breaking naming conventions, converting expressions into more complex forms, and declaring modifiers in the wrong order. Sorald also changed the original formatting of the code by causing indentation issues and relocating comments. Additionally, the repaired code was left with empty methods and unused method parameters. Most of these issues were introduced because Sorald’s current fixes only aim to address one specific SonarQube violation and do not account for side effects. When injecting code into the original codebase, Sorald has no awareness of the coding standards followed in the original code.\\
\textbf{Implication:} Although Sorald is able to fix specific SonarQube violations, it has no awareness that these fixes can lead to other issues. This highlights a key limitation in Sorald’s repair mechanism: the absence of semantic and stylistic understanding. 
Evidence here points to the need for APR evaluation to go beyond checks that focus on cleared fixes, as is proposed in this study.

\subsection{Evaluating Sorald’s Impact on Code Semantics (RQ3)}
Our outcomes showed that 1962 out of 8212 tests failed on the Sorald-repaired code, where 61 files were no longer compilable after applying Sorald. By analyzing the reasons for the compile errors, we observed the way partial fixes were applied, where Sorald repaired issues without understanding the context. Most importantly, changes made by Sorald altered code functionality. Another issue observed was that Sorald removed the entire implementation in some original files instead of repairing them. While this was previously evaluated as a suitable repair strategy when genetic improvement was applied to fix code violations\cite{lic2022,lic22022}, it raises serious questions for code quality. Since Sorald removed blocks of code from the original files in some cases, violations were removed simply because there was no code left. Although Sorald shows a high fix rate when analyzed with SonarQube (for RQ1), this does not necessarily mean that the issues were fixed correctly.\\
\textbf{Implication:} Our findings imply that relying solely on SonarQube's violation reduction as an indicator of successful repair is misleading, as Sorald's fixes may degrade program correctness or simply remove problematic code, with the opposite effect on code quality.

\subsection{Evaluating Sorald's Impact on Code Structure (RQ4)}
Outcomes showed that Sorald's repaired code had notable increases in certain code metrics---LCOM1, WMC, and LOC. LCOM1, an indicator of cohesion, measures how related the methods in a class are based on shared instance variables. A higher LCOM1 means the methods are more disjointed. This suggests that, after applying Sorald, the class loses relatedness. WMC (Weighted Methods per Class), which measures the complexity of a class, was also significantly increased after repairs. This shows that Sorald made the code more complex. When a class becomes complex, it is harder to test, understand, and maintain. The other highly affected metric is LOC (Lines Of Code), which is also related to code complexity. Sorald tends to add more code in the repairing process, reducing the readability and maintainability of the code. Moreover, Sorald also increased RFC (Response For a Class, complexity) and slightly decreased CBO (Coupling Between Object Classes). A slight reduction in CBO means Sorald has reduced the coupling between objects, increasing modularity and reusability.\\
\textbf{Implication:} While fixing rule violations, Sorald tends to degrade code quality by increasing complexity and reducing cohesion. This raises concerns about the maintainability, readability, and structural soundness of code after automated repairs.


\section{Threats to Validity}

\textbf{External Validity:} The evaluation was performed on code snippets extracted from Stack Overflow, which is a noisy dataset. This dataset does not represent well-structured Java codebases found on other platforms like GitHub. Moreover, our experiment is limited to 2393 Java files and Sorald. This does not represent all code on Stack Overflow, other programming languages, or APR tools.

\textbf{Construct Validity:} After applying Sorald there were line position shifts, affecting the accuracy of the fixed error count because the fixes are tracked using line numbers. This issue was addressed in this work through several checks as detailed in Section III. Manual evaluators were involved in checking the reasons for the new errors and test failures, with rigorous reliability checks performed. Sorald’s fixing capability is measured only by the number of fixes. We have also solely relied on SonarQube, which may produce false positives. Thus, these limitations may affect the depth of understanding on Sorald's fixing capability. That said, we have examined code repairs for contextual understanding through the analysis.

\section{Conclusion and Future Work}
In developing a comprehensive evaluation framework and demonstrating its utility, this study aimed to evaluate Sorald by investigating how many SonarQube violations are repaired for 30 fixable rules, Sorald’s tendency to introduce new errors, how Sorald affects the original semantics/functionality of code, and whether Sorald degrades code structure. While Sorald fixed 3423 errors out of 3529, its performance for two specific rules was below 50\%. We also found that Sorald generates a large number of code smells and several bugs during repairs, including breaking the coding standards of the original code, changing code formatting, causing precision loss in some arithmetic expressions, and creating race conditions. When analyzing how Sorald affects the original functionality, we noticed that 1962 out of 8212 test cases that passed on original code failed on Sorald-repaired code. The main reasons for these failures include some of Sorald’s fixes rendering Java files non-compilable, even though the SonarQube warnings were removed. Moreover, instead of repairing code, Sorald removed large chunks of code from some Java files. Some test failures were due to Sorald actually changing the code behavior --- the expected value and the actual value returned by functions were different. When investigating Sorald’s effect on code structure, Sorald notably increased LCOM1, LOC, and WMC of the code during repairs, degrading the original code structure by increasing complexity and reducing cohesion, which makes code harder to understand and maintain.

Our results highlight that Sorald can be considered a good tool when the fix is straightforward and has no effect on the rest of the code or other classes. However, Sorald lacks context-aware repair capability. It blindly introduces blocks of code or alters code just to fix a particular SonarQube violation, without considering the aftereffects. Additionally, some fixes are partial and lead to new violations, which indicates that Sorald’s current fixes should be improved before being applied to real-world code. We recommend manual review of Sorald's fixes and strict testing to detect any new or remaining deficiencies in code functionality. Moreover, from the original dataset, we removed 5,435 files because they did not violate any Sorald-fixable rules. It will be interesting to explore how Sorald behaves on fault-free code, where it is expected to ignore such files. While we concede a few threats to the study, the strategy followed to evaluate Sorald---looking not only at its fixing rate but also at how it affects code semantics, the likelihood of generating new errors, and how it affects the code structure---may serve as a benchmark for future research on APR tools' evaluation. To understand this behavior further, these tools should be evaluated on different datasets, our next step.
\balance
\section*{Acknowledgment}
This research is funded by New Zealand Ministry of Business, Innovation and Employment (MBIE) Smart Idea Award UOO2456.

\bibliographystyle{IEEEtran}
\bibliography{references.bib}

\begin{thebibliography}{10}
\providecommand{\url}[1]{#1}
\csname url@samestyle\endcsname
\providecommand{\newblock}{\relax}
\providecommand{\bibinfo}[2]{#2}
\providecommand{\BIBentrySTDinterwordspacing}{\spaceskip=0pt\relax}
\providecommand{\BIBentryALTinterwordstretchfactor}{4}
\providecommand{\BIBentryALTinterwordspacing}{\spaceskip=\fontdimen2\font plus
\BIBentryALTinterwordstretchfactor\fontdimen3\font minus
  \fontdimen4\font\relax}
\providecommand{\BIBforeignlanguage}[2]{{%
\expandafter\ifx\csname l@#1\endcsname\relax
\typeout{** WARNING: IEEEtran.bst: No hyphenation pattern has been}%
\typeout{** loaded for the language `#1'. Using the pattern for}%
\typeout{** the default language instead.}%
\else
\language=\csname l@#1\endcsname
\fi
#2}}
\providecommand{\BIBdecl}{\relax}
\BIBdecl

\bibitem{hamer2024just}
S.~Hamer, M.~d’Amorim, and L.~Williams, ``{Just another copy and paste?
  Comparing the security vulnerabilities of ChatGPT generated code and
  StackOverflow answers},'' in \emph{2024 IEEE Security and Privacy Workshops
  (SPW)}, 2024, pp. 87--94.

\bibitem{majdinasab2024assessing}
V.~Majdinasab, M.~J. Bishop, S.~Rasheed, A.~Moradidakhel, A.~Tahir, and
  F.~Khomh, ``{Assessing the Security of GitHub Copilot's Generated Code-A
  Targeted Replication Study},'' in \emph{2024 IEEE International Conference on
  Software Analysis, Evolution and Reengineering (SANER)}.\hskip 1em plus 0.5em
  minus 0.4em\relax IEEE, 2024, pp. 435--444.

\bibitem{zampetti2017open}
F.~Zampetti, S.~Scalabrino, R.~Oliveto, G.~Canfora, and M.~Di~Penta, ``{How
  open source projects use static code analysis tools in continuous integration
  pipelines},'' in \emph{2017 IEEE/ACM 14th International Conference on Mining
  Software Repositories (MSR)}.\hskip 1em plus 0.5em minus 0.4em\relax IEEE,
  2017, pp. 334--344.

\bibitem{imtiaz2019challenges}
N.~Imtiaz, A.~Rahman, E.~Farhana, and L.~Williams, ``{Challenges with
  responding to static analysis tool alerts},'' in \emph{2019 IEEE/ACM 16th
  International Conference on Mining Software Repositories (MSR)}.\hskip 1em
  plus 0.5em minus 0.4em\relax IEEE, 2019, pp. 245--249.

\bibitem{kim2013automatic}
D.~Kim, J.~Nam, J.~Song, and S.~Kim, ``Automatic patch generation learned from
  human-written patches,'' in \emph{35th International Conference on Software
  Engineering (ICSE)}.\hskip 1em plus 0.5em minus 0.4em\relax IEEE, 2013, pp.
  802--811.

\bibitem{yuan2018arja}
Y.~Yuan and W.~Banzhaf, ``{ARJA: Automated Repair of Java Programs via
  Multi-Objective Genetic Programming},'' \emph{IEEE Transactions on Software
  Engineering}, vol.~46, no.~10, pp. 1040--1067, 2020.

\bibitem{lutellier2020coconut}
T.~Lutellier, H.~V. Pham, L.~Pang, Y.~Li, M.~Wei, and L.~Tan, ``Coconut:
  combining context-aware neural translation models using ensemble for program
  repair,'' in \emph{29th ACM SIGSOFT International Symposium on Software
  Testing and Analysis (ISSTA)}, 2020, pp. 101--114.

\bibitem{wadhwa2023frustrated}
N.~Wadhwa, J.~Pradhan, A.~Sonwane, S.~P. Sahu, N.~Natarajan, A.~Kanade,
  S.~Parthasarathy, and S.~Rajamani, ``{Frustrated with code quality issues?
  llms can help!}'' \emph{arXiv preprint arXiv:2309.12938}, 2023.

\bibitem{etemadi2022sorald}
K.~Etemadi, N.~Harrand, S.~Larsén, H.~Adzemovic, H.~L. Phu, A.~Verma,
  F.~Madeiral, D.~Wikström, and M.~Monperrus, ``{Sorald: Automatic Patch
  Suggestions for SonarQube Static Analysis Violations},'' \emph{IEEE
  Transactions on Dependable and Secure Computing}, vol.~20, no.~4, pp.
  2794--2810, 2023.

\bibitem{yin2024thinkrepair}
X.~Yin, C.~Ni, S.~Wang, Z.~Li, L.~Zeng, and X.~Yang, ``{Thinkrepair:
  Self-directed automated program repair},'' in \emph{33rd ACM SIGSOFT
  International Symposium on Software Testing and Analysis (ISSTA)}, 2024, pp.
  1274--1286.

\bibitem{li2024hybrid}
F.~Li, J.~Jiang, J.~Sun, and H.~Zhang, ``{Hybrid automated program repair by
  combining large language models and program analysis},'' \emph{ACM
  Transactions on Software Engineering and Methodology}, 2024.

\bibitem{just2014defects4j}
R.~Just, D.~Jalali, and M.~D. Ernst, ``{Defects4J: A database of existing
  faults to enable controlled testing studies for Java programs},'' in
  \emph{International Symposium on Software Testing and Analysis (ISSTA)},
  2014, pp. 437--440.

\bibitem{lin2017quixbugs}
D.~Lin, J.~Koppel, A.~Chen, and A.~Solar-Lezama, ``{QuixBugs: A multi-lingual
  program repair benchmark set based on the Quixey Challenge},'' in
  \emph{Proceedings Companion of the ACM SIGPLAN International Conference on
  Systems, Programming, Languages, and Applications: Software for Humanity
  (SPLASH)}, 2017, pp. 55--56.

\bibitem{ndukwe2023have}
I.~G. Ndukwe, S.~A. Licorish, A.~Tahir, and S.~G. MacDonell, ``{How have views
  on software quality differed over time? Research and practice viewpoints},''
  \emph{Journal of Systems and Software}, vol. 195, p. 111524, 2023.

\bibitem{lotter2018code}
A.~Lotter, S.~A. Licorish, B.~T.~R. Savarimuthu, and S.~Meldrum, ``{Code reuse
  in stack overflow and popular open source java projects},'' in \emph{25th
  Australasian Software Engineering Conference (ASWEC)}.\hskip 1em plus 0.5em
  minus 0.4em\relax IEEE, 2018, pp. 141--150.

\bibitem{licorish2021contextual}
S.~A. Licorish and T.~Nishatharan, ``{Contextual profiling of stack overflow
  java code security vulnerabilities initial insights from a pilot study},'' in
  \emph{IEEE 21st International Conference on Software Quality, Reliability and
  Security Companion (QRS-C)}.\hskip 1em plus 0.5em minus 0.4em\relax IEEE,
  2021, pp. 1060--1068.

\bibitem{licorish2025comparing}
S.~A. Licorish, A.~Bajpai, C.~Arora, F.~Wang, and K.~Tantithamthavorn,
  ``{Comparing Human and LLM Generated Code: The Jury is Still Out!}''
  \emph{arXiv preprint arXiv:2501.16857}, 2025.

\bibitem{chong2024artificial}
C.~J. Chong, Z.~Yao, and I.~Neamtiu, ``{Artificial-Intelligence Generated Code
  Considered Harmful: A Road Map for Secure and High-Quality Code
  Generation},'' \emph{arXiv preprint arXiv:2409.19182}, 2024.

\bibitem{arora2024optimizing}
C.~Arora, A.~I. Sayeed, S.~Licorish, F.~Wang, and C.~Treude, ``{Optimizing
  large language model hyperparameters for code generation},'' \emph{arXiv
  preprint arXiv:2408.10577}, 2024.

\bibitem{gomes2009overview}
I.~Gomes, P.~Morgado, T.~Gomes, and R.~Moreira, ``{An overview on the static
  code analysis approach in software development},'' \emph{Faculdade de
  Engenharia da Universidade do Porto, Portugal}, vol.~16, 2009.

\bibitem{10371505}
S.~Nocera, S.~Romano, R.~Francese, R.~Burlon, and G.~Scanniello, ``{Managing
  Vulnerabilities in Software Projects: the Case of NTT Data},'' in \emph{2023
  49th Euromicro Conference on Software Engineering and Advanced Applications
  (SEAA)}, 2023, pp. 247--253.

\bibitem{10.1145/3544902.3546233}
\BIBentryALTinterwordspacing
S.~Romano, F.~Zampetti, M.~T. Baldassarre, M.~Di~Penta, and G.~Scanniello,
  ``{Do Static Analysis Tools Affect Software Quality when Using Test-driven
  Development?}'' in \emph{Proceedings of the 16th ACM / IEEE International
  Symposium on Empirical Software Engineering and Measurement}, ser. ESEM
  '22.\hskip 1em plus 0.5em minus 0.4em\relax New York, NY, USA: Association
  for Computing Machinery, 2022, p. 80–91. [Online]. Available:
  \url{https://doi.org/10.1145/3544902.3546233}
\BIBentrySTDinterwordspacing

\bibitem{wadhams2024automating}
Z.~Wadhams, A.~M. Reinhold, and C.~Izurieta, ``{Automating Static Code Analysis
  Through CI/CD Pipeline Integration},'' in \emph{2024 IEEE International
  Conference on Software Analysis, Evolution and Reengineering - Companion
  (SANER-C)}, 2024, pp. 119--125.

\bibitem{mechtaev2015directfix}
S.~Mechtaev, J.~Yi, and A.~Roychoudhury, ``{DirectFix: Looking for Simple
  Program Repairs},'' in \emph{2015 IEEE/ACM 37th IEEE International Conference
  on Software Engineering}, vol.~1, 2015, pp. 448--458.

\bibitem{bouzenia2403repairagent}
I.~Bouzenia, P.~Devanbu, and M.~Pradel, ``{Repairagent: an autonomous,
  llm-based agent for program repair.(2024)},'' \emph{arXiv preprint
  arXiv:2403.17134}.

\bibitem{lic2022}
S.~A. Licorish and M.~Wagner, ``{Combining GIN and PMD for code
  improvements},'' in \emph{Genetic and Evolutionary Computation Conference
  Companion (GECCO)}.\hskip 1em plus 0.5em minus 0.4em\relax ACM, 2022, p.
  790–793.

\bibitem{meldrum2020understanding}
S.~Meldrum, S.~A. Licorish, C.~A. Owen, and B.~T.~R. Savarimuthu,
  ``{Understanding stack overflow code quality: A recommendation of caution},''
  \emph{Science of Computer Programming}, vol. 199, p. 102516, 2020.

\bibitem{jiang2008comparing}
Y.~Jiang, B.~Cuki, T.~Menzies, and N.~Bartlow, ``{Comparing design and code
  metrics for software quality prediction},'' in \emph{4th International
  Workshop on Predictor Models in Software Engineering (PROMISE)}, 2008, pp.
  11--18.

\bibitem{liyanage_2025_16826817}
\BIBentryALTinterwordspacing
S.~Liyanage, S.~Licorish, M.~Wagner, and S.~MacDonell, ``{On the need to
  perform comprehensive evaluations of automated program repair benchmarks:
  Sorald case study - Replication Package },'' Aug. 2025. [Online]. Available:
  \url{https://doi.org/10.5281/zenodo.16826817}
\BIBentrySTDinterwordspacing

\bibitem{nguyen2023software}
T.~Nguyen, Y.~Di, J.~Lee, M.~Chen, and T.~Zhang, ``{Software Entity Recognition
  with Noise-Robust Learning},'' in \emph{2023 38th IEEE/ACM International
  Conference on Automated Software Engineering (ASE)}, 2023, pp. 484--496.

\bibitem{jimenez2016vulnerability}
M.~Jimenez, M.~Papadakis, and Y.~L. Traon, ``{Vulnerability Prediction Models:
  A Case Study on the Linux Kernel},'' in \emph{2016 IEEE 16th International
  Working Conference on Source Code Analysis and Manipulation (SCAM)}, 2016,
  pp. 1--10.

\bibitem{fraser2011evosuite}
G.~Fraser and A.~Arcuri, ``{Evosuite: automatic test suite generation for
  object-oriented software},'' in \emph{19th ACM SIGSOFT Symposium and the 13th
  European Conference on Foundations of Software Engineering (ESEC/FSE)}, 2011,
  pp. 416--419.

\bibitem{rebro2023source}
\BIBentryALTinterwordspacing
D.~A. Rebro, S.~Chren, and B.~Rossi, ``{Source Code Metrics for Software
  Defects Prediction},'' in \emph{Proceedings of the 38th ACM/SIGAPP Symposium
  on Applied Computing}, ser. SAC '23.\hskip 1em plus 0.5em minus 0.4em\relax
  New York, NY, USA: Association for Computing Machinery, 2023, p. 1469–1472.
  [Online]. Available: \url{https://doi.org/10.1145/3555776.3577809}
\BIBentrySTDinterwordspacing

\bibitem{gyimothy2005empirical}
T.~Gyim{\'o}thy, R.~Ferenc, and I.~Siket, ``{Empirical validation of
  object-oriented metrics on open source software for fault prediction},''
  \emph{IEEE Transactions on Software Engineering}, vol.~31, no.~10, pp.
  897--910, 2005.

\bibitem{chidamber1994metrics}
S.~R. Chidamber and C.~F. Kemerer, ``{A metrics suite for object oriented
  design},'' \emph{IEEE Transactions on Software Engineering}, vol.~20, no.~6,
  pp. 476--493, 1994.

\bibitem{aniche-ck}
M.~Aniche, \emph{Java code metrics calculator (CK)}, 2015, available in
  https://github.com/mauricioaniche/ck/.

\bibitem{lic22022}
S.~A. Licorish and M.~Wagner, ``{Dissecting copy/delete/replace/swap mutations:
  insights from a GIN case study},'' in \emph{Genetic and Evolutionary
  Computation Conference Companion (GECCO)}.\hskip 1em plus 0.5em minus
  0.4em\relax ACM, 2022, p. 1940–1945.

\end{thebibliography}

\vspace{12pt}

\end{document}